\documentclass[journal]{IEEEtran}
\usepackage{amsmath,amssymb}
\usepackage{graphicx,booktabs,cite,url,color}
\usepackage{algorithm,algpseudocode}

\begin{document}
\title{\huge Quantum Optimization Algorithm for Pinching-Antenna Activation with Max-Min Fairness}

\author{Soumyadip Paul, \IEEEmembership{Student Member, IEEE} and Neel Kanth Kundu, \IEEEmembership{Member, IEEE}
\thanks{The work of Soumyadip Paul was supported in part by the IndiaAI Fellowship (App. ID.: PHD251015004118) under the IndiaAI Mission of the Ministry of Electronics and Information Technology (MeitY), Government of India. The work of Neel Kanth Kundu was supported in part by the INSPIRE Faculty Fellowship (Reg. No.: IFA22-ENG 344), ANRF Prime Minister Early Career Research Grant (ANRF/ECRG/2024/000324/ENS), and the New Faculty Seed Grant from the Indian Institute of Technology Delhi. This work was also supported by the University of Melbourne through the establishment of an IBM Quantum Network Hub.}%
\thanks{Soumyadip Paul is with the Center for Applied Research in Electronics (CARE), Indian Institute of Technology Delhi, New Delhi-110016, India (e-mail: crz258451@care.iitd.ac.in).}%
\thanks{Neel Kanth Kundu is with CARE and the Bharti School of Telecommunication Technology and Management, Indian Institute of Technology Delhi, New Delhi-110016, India (e-mail: neelkanth@iitd.ac.in). He is also an honorary fellow at the Department of Electrical and Electronic Engineering, University of Melbourne, Melbourne, VIC-3010, Australia.}}

\maketitle

\begin{abstract}
In this letter, we propose a training-free quantum approximate optimization algorithm with a linear-ramp schedule (LR-QAOA) for antenna activation in pinching-antenna systems (PASS). Since all active pinching antennas (PAs) share one feed and split its power, maximizing the signal-to-noise ratio (SNR) of the weakest user becomes a combinatorial max-min problem that must be re-solved whenever the users move. We formulate the problem as a two-local Ising Hamiltonian with one qubit per PA and no penalty terms, employing a weighted-sum surrogate to prioritize weaker users. 
Instead of training the circuit with a classical optimizer, the proposed LR-QAOA method tunes its parameters via a linear ramp; the measured patterns are ranked by the exact max-min objective, and the best ones are refined with a short local search using the same number of evaluated patterns. Monte Carlo simulations show that the proposed LR-QAOA method outperforms greedy activation, projection-guided activation, and semidefinite relaxation-based classical baseline methods when all methods test the same number of PA on/off patterns. Experiments on real IBM Quantum hardware further show that the execution time remains nearly constant as the number of PAs grows, whereas exhaustive search grows exponentially. However, gate noise in the NISQ era restricts near-optimal hardware solutions to small numbers of antennas only.
\end{abstract}

\begin{IEEEkeywords}
Pinching antennas, antenna activation, QAOA, linear ramp, Ising model, max-min fairness.
\end{IEEEkeywords}

\section{Introduction}
\label{sec:intro}
\IEEEPARstart{P}{inching-antenna} systems (PASS) carry a radio-frequency signal along a low-loss dielectric waveguide and radiate it only where a small dielectric ``pinch'' touches the guide \cite{fukuda2022}. Each pinch acts as an antenna, and since the waveguide can extend over many meters, the radiating points can be placed close to the users. PASS thus converts the problem of large-scale path loss into one of selecting good positions, establishing strong line-of-sight links with very simple hardware \cite{liu2025tut}.

Since moving the pinches mechanically is slow, low-complexity deployments pre-install the pinching antennas (PAs) at fixed positions and only switch them on or off \cite{wang2026,cao2026,vss2026}. However, this simplification comes at an algorithmic price. Because all active PAs share a single feed, their signals add coherently while the transmit power is split among them. Consequently, a new PA adds a path but weakens every other one, and a PA that adds in phase at one user may cancel at another. For $N$ PAs, the problem is combinatorial, with $2^N-1$ candidate patterns, and must be solved again whenever the users move.

When one PA pattern serves several users, fairness matters. Maximizing the signal-to-noise ratio (SNR) of the weakest user leads to a max-min fractional 0--1 program \cite{vss2026}, which is a quadratic fractional problem even for a single user \cite{karag2025}. Existing approaches trade optimality for speed, e.g., game-theoretic activation \cite{wang2026}, projection-guided greedy activation (PGGA) \cite{cao2026}, deep learning \cite{karag2025}, and trellis-based Viterbi state selection (VSS)\cite{vss2026}. Semidefinite relaxation (SDR) \cite{luo2010} remains the classical general-purpose tool for such binary quadratic problems. As shown in this work, greedy methods tend to stop too early, since the power split makes almost every addition appear harmful, whereas exhaustive search becomes impractical for real-time operation beyond about 20 PAs.

Quantum optimization offers a different route. The quantum approximate optimization algorithm (QAOA) \cite{farhi2014,blekos2024} prepares a parameterized quantum state that favors low-energy solutions of an Ising Hamiltonian and samples from it, which makes it a natural fit for binary optimization problems. Prior works have investigated quantum optimization for MIMO detection \cite{cui2022,kim2019}, and QAOA for RIS phase optimization \cite{paul2026lr} and MIMO-QAM data detection \cite{paul2026ws}. Despite these works, several research gaps remain. First, to the best of our knowledge, PA activation has not yet been formulated as an Ising problem, and such a formulation is not immediate, since the max-min objective contains both a ratio and a minimum, neither of which is quadratic. Second, standard QAOA tunes $2p$ circuit angles for every instance with a classical optimizer, which is too slow when every channel realization is a new instance. Linear-ramp QAOA (LR-QAOA) avoids this by fixing all angles through a single scale parameter that transfers across instances \cite{montanez2025}, but has not been applied to PA activation problem yet. Third, gate noise, which grows with circuit size, limits current hardware, so both the execution time and the achieved quality must be evaluated on real quantum devices. This work addresses these three gaps through the following contributions:
\begin{itemize}
\item We formulate max-min PA activation as a two-local Ising Hamiltonian with one qubit per PA and no penalty terms, and analyze how well its weighted-sum surrogate ranks the true optimum.
\item We propose a training-free LR-QAOA framework with post-selection on the exact objective and a short local refinement. We compare the proposed LR-QAOA framework with greedy activation, PGGA, and SDR-based classical baseline methods.
\item We validate the proposed QAOA approach on IBM Quantum hardware with up to 120 qubits and report both the quantum processing unit (QPU) time and the solution quality actually achieved.
\end{itemize}

Together, these contributions show that LR-QAOA finds better activation patterns than greedy, PGGA~\cite{cao2026}, and SDR~\cite{luo2010} in the multi-user case. Moreover, its time complexity scales more favorably than that of exhaustive search, which makes LR-QAOA a scalable candidate for PASS as quantum hardware improves, since its complexity grows polynomially with the number of PAs and linearly with the number of users.

The remainder of this work is organized as follows. Section~\ref{sec:system} presents the system model and the Ising formulation. Section~\ref{sec:hybrid} presents the LR-QAOA algorithm, and Section~\ref{sec:results} presents the simulation and hardware results, while Section~\ref{sec:conclusion} concludes the paper.

{\em Notations}: Matrices and vectors are denoted by boldface uppercase ($\mathbf A$) and lowercase ($\mathbf a$) letters, respectively. The transpose and the conjugate transpose of $\mathbf A$ are denoted by $\mathbf A^T$ and $\mathbf A^H$, respectively. $\mathrm{Re}(\cdot)$ denotes the real part, $|\cdot|$ the absolute value, $\|\cdot\|$ the Euclidean norm, and $\mathbf 1$ the all-ones vector. A complex Gaussian random variable with mean $\theta$ and variance $\sigma^2$ is denoted by $\mathcal{CN}(\theta,\sigma^2)$. $X_n$ and $Z_n$ denote the Pauli-$X$ and Pauli-$Z$ operators acting on qubit $n$, $|0\rangle$ and $|1\rangle$ the computational basis states, $|+\rangle=(|0\rangle+|1\rangle)/\sqrt2$, and $(\cdot)^{\otimes N}$ the $N$-fold tensor product. The single-qubit rotation gates are $R_X(\theta)=e^{-i\theta X/2}$ and $R_Z(\theta)=e^{-i\theta Z/2}$, the two-qubit rotation is $R_{ZZ}(\theta)=e^{-i\theta Z\otimes Z/2}$, and CNOT denotes the controlled-NOT gate. 
\begin{figure}[t]
\centering
\includegraphics[width=0.9\columnwidth]{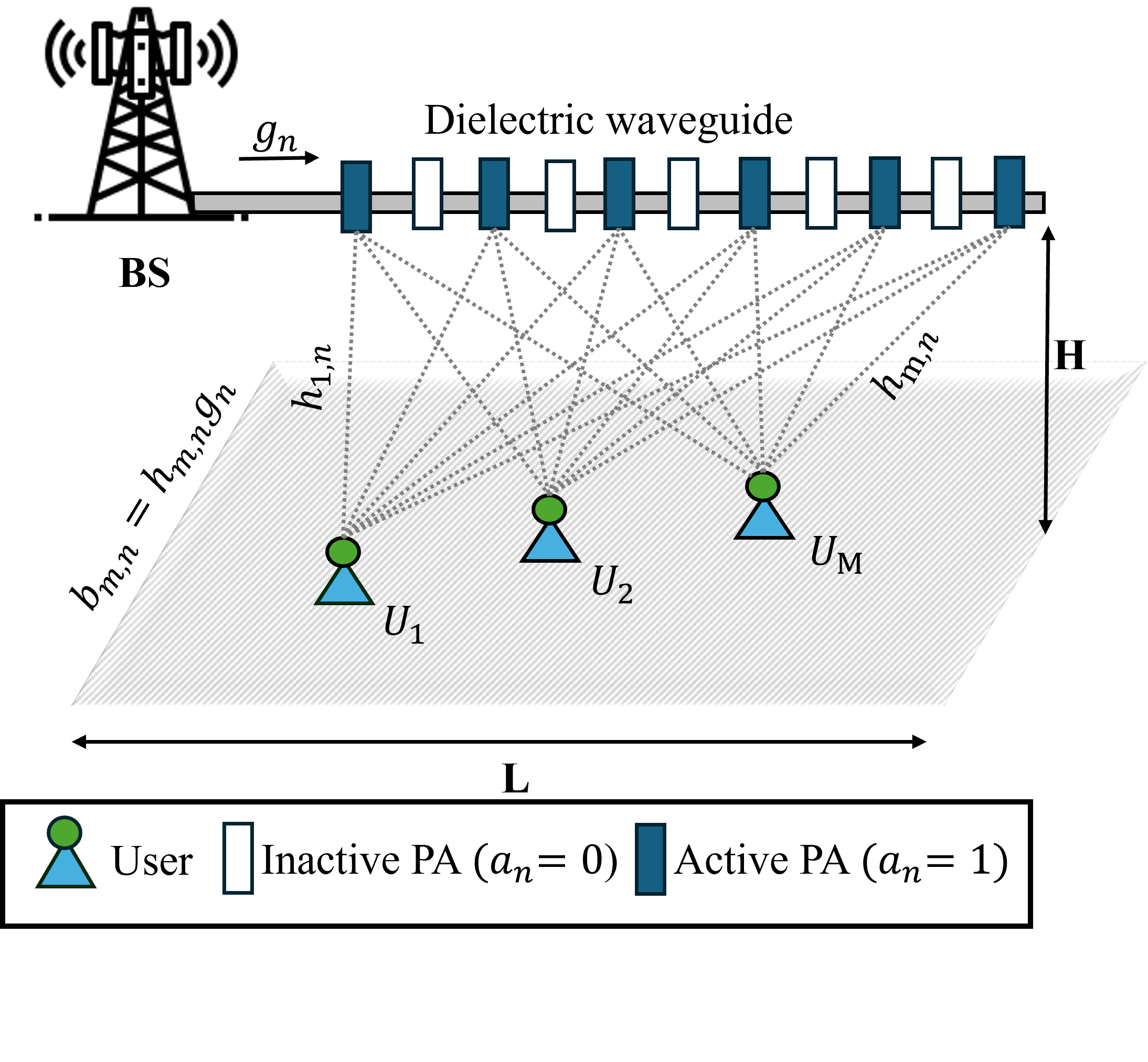}
\caption{PASS with a waveguide of length $L$ at height $H$ carrying $N$ pre-installed PAs; one activation pattern $\mathbf a$ serves $M$ users.}
\label{fig:sys}
\end{figure}

\section{System Model and Ising Formulation}
\label{sec:system}
\subsection{PASS Model}
We consider the PASS model of \cite{vss2026}, as shown in Fig.~\ref{fig:sys}. A waveguide of length $L$ is deployed along the $x$-axis at height $H$ and is fed at $\boldsymbol\psi_0=(-L/2,0,H)$. The $N$ PAs are located at $\boldsymbol\psi_n=\big(-\tfrac L2+\tfrac{(2n-1)L}{2N},0,H\big)$, and $M$ single-antenna users are located at $\boldsymbol\psi_m=(x_m,y_m,0)$, where $x_m$ and $y_m$ are uniformly distributed in $[-L/2,L/2]$ and are assumed to be known at the base station. Neglecting in-waveguide loss, the effective channel $b_{m,n}=h_{m,n}g_n$ from PA $n$ to user $m$ is given by
\begin{equation}
b_{m,n}=\underbrace{\frac{e^{-j\frac{2\pi}{\lambda}\|\boldsymbol\psi_m-\boldsymbol\psi_n\|}}{\|\boldsymbol\psi_m-\boldsymbol\psi_n\|}}_{h_{m,n}}\,\underbrace{e^{-j\frac{2\pi}{\lambda_g}\|\boldsymbol\psi_n-\boldsymbol\psi_0\|}}_{g_n},
\label{eq:channel}
\end{equation}
where $\lambda_g=\lambda/n_{\rm eff}$ denotes the guided wavelength, and $\mathbf b_m=[b_{m,1},\dots,b_{m,N}]^T$ denotes the channel vector of the $m$-th user. An activation vector $\mathbf a\in\{0,1\}^N$ switches on $\mathbf 1^T\mathbf a$ PAs, which share the transmit power $P$ equally. Under time-division access, the SNR of user $m$ is given by

\begin{equation}
\gamma_m(\mathbf a)=\frac{P\eta}{\sigma_w^2}\,\frac{|\mathbf a^T\mathbf b_m|^2}{\mathbf 1^T\mathbf a},
\end{equation}
where $\eta=c^2/(16\pi^2f_c^2)$ is the free-space path-loss scaling factor, $c$ denotes the speed of light, $f_c$ the carrier frequency, $P$ the transmit power, and $\sigma_w^2$ the variance of the additive white Gaussian noise $w\sim\mathcal{CN}(0,\sigma_w^2)$. The corresponding achievable rate is $\log_2(1+\gamma_m)$. To ensure fairness across users, the max-min activation problem is given by \cite{vss2026}
\begin{equation}
\max_{\mathbf a\in\{0,1\}^N\setminus\{\mathbf 0\}}\ \Gamma(\mathbf a),\qquad \Gamma(\mathbf a)=\min_{m}\frac{|\mathbf a^T\mathbf b_m|^2}{\mathbf 1^T\mathbf a}.
\label{eq:P}
\end{equation}

\subsection{Ising Formulation of the Max-Min Problem}
\label{sec:ising}
 To be solved by QAOA, the optimization problem must be expressed as a quadratic function of the binary activation variables, i.e., as an Ising Hamiltonian. However, the problem in \eqref{eq:P}, contains a ratio and a minimum over the users, neither of which is quadratic. Therefore, we  replace it by a quadratic surrogate, which we build in three steps. First, note that for binary $\mathbf a$, the numerator is a quadratic form given by $|\mathbf a^T\mathbf b_m|^2=\mathbf a^T\mathbf R_m\mathbf a$, where $\mathbf R_m=\mathrm{Re}(\mathbf b_m\mathbf b_m^H)$. The off-diagonal entries of $\mathbf R_m$ are positive for PA pairs that arrive in phase at user $m$ and negative for pairs that cancel. Second, following Dinkelbach's method \cite{dinkelbach1967}, $\Gamma(\mathbf a)\ge t$ holds if and only if $\mathbf a^T\mathbf R_m\mathbf a-t\,\mathbf 1^T\mathbf a\ge0$ for every user, and hence the ratio becomes a price on each active PA. Instead of iterating on $t$, we adopt a fixed price $\mu$. Third, we replace the minimum by a weighted sum with weights $w_m=\|\mathbf b_m\|^{-2\nu}$, which assigns more weight to users with weak channels. Combining these steps yields the surrogate objective
\begin{equation}
F(\mathbf a)=\mathbf a^T\mathbf R\,\mathbf a-\mu\,\mathbf 1^T\mathbf a,\qquad \mathbf R=\sum_mw_m\mathbf R_m,
\label{eq:F}
\end{equation}
where $\mu=\kappa\max_nR_{nn}$ is a fraction of the weighted power of the best single PA. Since $a_n^2=a_n$, maximizing $F$ is equivalent to minimizing $C(\mathbf a)=\sum_n(\mu-R_{nn})a_n-2\sum_{n<n'}R_{nn'}a_na_{n'}$. Substituting $a_n=(1-z_n)/2$ and replacing $z_n$ by the Pauli operator $Z_n$ yields, up to a constant, the Ising Hamiltonian
\begin{equation}
H_C=\sum_{n}h_nZ_n+\sum_{n<n'}J_{nn'}Z_nZ_{n'},
\label{eq:HC}
\end{equation}
with local fields and pairwise couplings given by
\begin{equation}
h_n=\frac{R_{nn}-\mu}{2}+\frac12\sum_{n'\ne n}R_{nn'},\qquad J_{nn'}=-\frac{R_{nn'}}{2},
\label{eq:hJ}
\end{equation}
where $|1\rangle_n$ indicates that PA $n$ is active. Since the problem is unconstrained, $H_C$ requires no penalty terms, and every measured bitstring except the all-zero one corresponds to a valid pattern. 

The surrogate $F(\mathbf a)$ is bounded below by an affine function of $\Gamma(\mathbf a)$ as
\begin{equation}
F(\mathbf a)\ge\Big(\big(\textstyle\sum_mw_m\big)\Gamma(\mathbf a)-\mu\Big)\mathbf 1^T\mathbf a,
\label{eq:bounds}
\end{equation}
since every user's gain is at least that of the worst user, with equality for a single user. Hence, a pattern with a good worst user always has a large $F$, i.e., a low energy of $H_C$, so the max-min optimum remains among the low-energy patterns that QAOA favors, and post-selection on $\Gamma$ picks it among the sampled patterns. Since a large $F$ does not guarantee a large $\Gamma$, this bound loosens as users are added, which is why we add local refinement in Section~\ref{sec:hybrid}. Finally, $0 \leq \kappa \leq 1$ sets the price of each active PA, such that $\kappa=0$ activates too many PAs and $\kappa=1$ activates very few or none.

\begin{figure}[t]
\centering
\includegraphics[width=1\columnwidth]{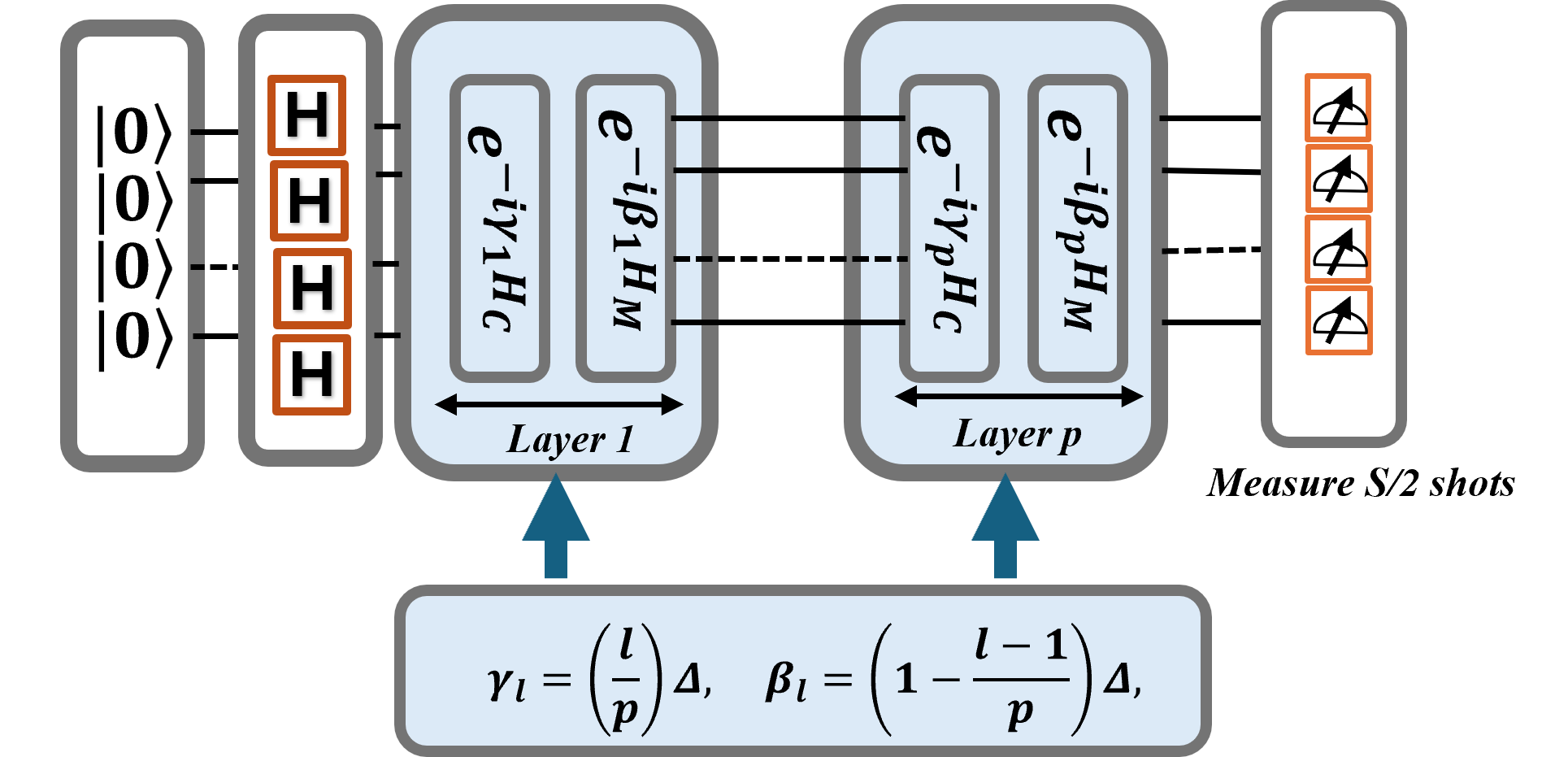}
\caption{LR-QAOA circuit with $p$ linear-ramp layers; measured patterns are post-selected on $\Gamma$ and refined by local search.}
\label{fig:circ}
\end{figure}

\section{LR-QAOA With Local Refinement}
\label{sec:hybrid}
 Starting from the initial state $|+\rangle^{\otimes N}$, which is the ground state of the mixer Hamiltonian $H_M=-\sum_nX_n$ and assigns equal probability $2^{-N}$ to every activation pattern, a depth-$p$ circuit prepares
\begin{equation}
|\psi_p\rangle=e^{-i\beta_p H_M}e^{-i\gamma_p H_C}\cdots e^{-i\beta_1 H_M}e^{-i\gamma_1 H_C}\,|+\rangle^{\otimes N}.
\end{equation}
Since all terms of $H_C$ in \eqref{eq:HC} are diagonal in the computational basis and therefore commute, the cost unitary factorizes exactly into single- and two-qubit rotations,
\begin{equation}
e^{-i\gamma H_C}=\prod_{n}R_Z(2\gamma h_n)\prod_{n<n'}R_{ZZ}(2\gamma J_{nn'}),
\label{eq:UC}
\end{equation}
where $R_{ZZ}$ acts on qubits $n$ and $n'$. The cost layer imprints on each pattern a phase proportional to its cost, so that, for the computational basis state $|\mathbf a\rangle$, $e^{-i\gamma H_C}|\mathbf a\rangle=e^{-i\gamma E(\mathbf a)}|\mathbf a\rangle$, where $E(\mathbf a)$ denotes the energy of $\mathbf a$ under $H_C$. Similarly, the mixer factorizes into independent single-qubit rotations $e^{-i\beta H_M}=\prod_nR_X(-2\beta)$, which let the amplitudes of patterns that differ in one PA interfere. Through this interplay of phases and interference, the amplitudes of low-energy patterns grow, whereas those of high-energy patterns are suppressed. As depicted in Fig.~\ref{fig:circ}, each cost layer thus uses one $R_Z$ gate per PA and one CNOT--$R_Z$--CNOT block per coupling, whereas each mixer layer uses $N$ $R_X$ rotations. Since $\mathbf R$ is generally dense, every pair of PAs is coupled, and one layer contains $N(N-1)/2$ two-qubit blocks.

\begin{algorithm}[t]
\caption{LR-QAOA for Max-Min PA Activation}
\label{alg}
\small
\begin{algorithmic}[1]
\Require $\{\mathbf b_m\}$, depth $p$, scale $\Delta$, factors $\kappa$, $\nu$, number of evaluated patterns $S$
\Statex \textit{Phase 1: Hamiltonian and schedule (classical)}
\State $\mathbf R\gets\sum_m\|\mathbf b_m\|^{-2\nu}\mathrm{Re}(\mathbf b_m\mathbf b_m^H)$, $\mu\gets\kappa\max_nR_{nn}$
\State Compute $h_n$, $J_{nn'}$ by \eqref{eq:hJ} and $\gamma_\ell$, $\beta_\ell$ by \eqref{eq:ramp}
\Statex \textit{Phase 2: Sampling (quantum)}
\State From $|+\rangle^{\otimes N}$, apply $p$ layers of cost ($R_Z$, $R_{ZZ}$) and mixer ($R_X$)
\State Measure $S/2$ shots; $\mathcal A\gets$ distinct nonzero patterns
\Statex \textit{Phase 3: Post-selection and refinement (classical)}
\State Sort $\mathcal A$ by decreasing $\Gamma$; $\mathbf a^\star\gets$ best of $\mathcal A$
\For{each $\mathbf a\in\mathcal A$, until $S/2$ more evaluations are used}
\Repeat\ flip PAs of $\mathbf a$ one at a time in random order;
\Statex \hspace{\algorithmicindent}\hspace{\algorithmicindent}accept the first flip that raises $\Gamma$
\Until{no flip improves $\Gamma$}
\State $\mathbf a^\star\gets\arg\max\{\Gamma(\mathbf a^\star),\Gamma(\mathbf a)\}$
\EndFor
\State \Return $\mathbf a^\star$
\end{algorithmic}
\end{algorithm}

Measuring $|\psi_p\rangle$ in the computational basis returns pattern $\mathbf a$ with probability $|\langle\mathbf a|\psi_p\rangle|^2$, where outcome $|1\rangle_n$ is read as $a_n=1$. QAOA is therefore used here as a sampler rather than as an estimator of $\langle H_C\rangle$: we only require that good patterns appear among the measured ones, not that the most frequent pattern be optimal. This viewpoint matters because, as shown in Section~\ref{sec:sim}, the ground state of $H_C$ is often not the max-min optimum.

Standard QAOA tunes the $2p$ angles $\{\gamma_\ell,\beta_\ell\}$ for each instance with a classical optimizer, which requires many repeated circuit executions per channel realization. Rather than training the angles, we fix them by a linear-ramp schedule, given by \cite{montanez2025,sack2021}
\begin{equation}
\gamma_\ell=\frac{\ell}{p}\Delta,\qquad\beta_\ell=\Big(1-\frac{\ell-1}{p}\Big)\Delta,
\label{eq:ramp}
\end{equation}
where the single scale parameter $\Delta$ is selected once offline and applied to all instances. The schedule in \eqref{eq:ramp} mimics a first-order Trotterization of the annealing path $H(s)=(1-s)H_M+sH_C$, $s\in[0,1]$, with $\Delta$ acting as the time step: the cost angles increase while the mixer angles decrease, so the circuit gradually shifts weight from $H_M$ to $H_C$. A small $\Delta$ keeps the state close to $|+\rangle^{\otimes N}$, so that the samples approach uniform sampling, whereas a very large $\Delta$ increases the discretization error. Within a broad intermediate range, the samples concentrate on low-energy patterns without any per-instance training, which makes the schedule suitable when the users move and a new instance arises in every channel realization.

As outlined in Algorithm~\ref{alg}, the $S/2$ measured patterns are scored with the exact objective $\Gamma$ rather than the surrogate, which corrects for the mismatch between $F$ and $\Gamma$ discussed in Section~\ref{sec:ising}. The remaining $S/2$ evaluations are spent on a single-flip local search started from the best measured patterns, so exactly $S$ patterns are evaluated in total, as for every sampling baseline. In the remainder of this work, ``LR-QAOA'' refers to this complete procedure.

\emph{Complexity Analysis:} The LR-QAOA circuit requires $N$ qubits and $O(pN^2)$ gate operations, or a depth of $O(pN)$ with a swap network \cite{ogorman2019}. Phases~1 and~3 require $O(N^2M+SNM)$ operations, and no classical optimization loop is involved. Moreover, since one evaluation of $\Gamma$ takes about 5\,$\mu$s on a CPU, whereas one shot on current hardware takes 1.4--7.8\,ms including overheads, replacing half of the shots by local search halves the number of shots and thus reduces the QPU time. In contrast, exhaustive search requires $O(2^NM)$ operations with Gray-code updates.

\section{Numerical Results}
\label{sec:results}
\subsection{Simulation Results}
\label{sec:sim}
We consider a simulation scenario by adopting the parameters of \cite{vss2026}, i.e., $H=3$\,m, $L=50$\,m, $f_c=28$\,GHz, $n_{\rm eff}=1.4$, a noise power of $-90$\,dBm, and $P=10$\,dBm, and use $M=3$, $p=5$, and $\Delta=1.75$ unless stated otherwise. The scale $\Delta$ is selected once offline on 60 held-out instances with $N=12$ and kept fixed for all other instances. LR-QAOA is compared with greedy activation, PGGA~\cite{cao2026}, and SDR with Gaussian randomization~\cite{luo2010} using the same number of evaluated patterns $S$. The surrogate does not always place the optimum at the ground state of $H_C$, especially when several users are served. Still, the optimum usually lies among the lowest-energy patterns, so post-selection on $\Gamma$ picks it up from the samples. A small $\kappa$ favors many active PAs, which the power split penalizes. In practice, the choice of $\kappa$ matters little, since the loss is almost the same for any $\kappa$ between 0.7 and 1. We therefore fix $\kappa=0.75$ offline, just like $\Delta$, and use $\nu=1$.

\begin{figure*}[!t]
\centering
\begin{tabular}{@{}c@{\hspace{2mm}}c@{\hspace{2mm}}c@{}}
\includegraphics[width=0.32\textwidth]{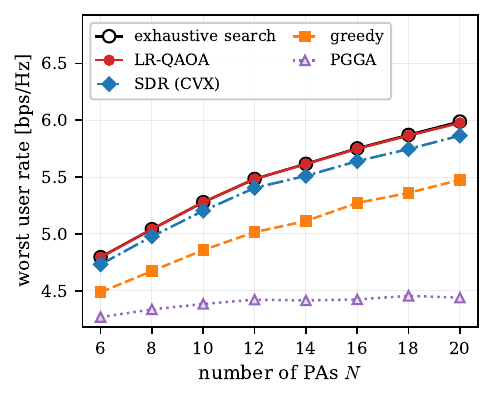} &
\includegraphics[width=0.32\textwidth]{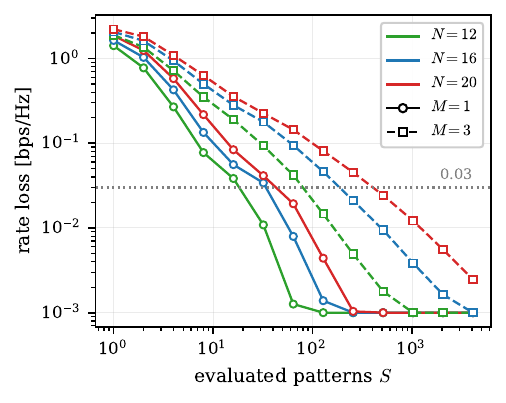} &
\includegraphics[width=0.32\textwidth]{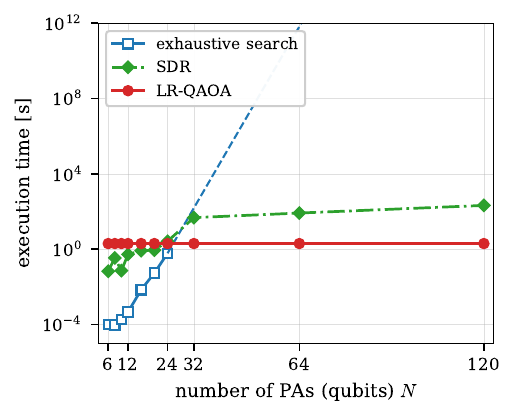}\\[-1mm]
\footnotesize (a) & \footnotesize (b) & \footnotesize (c)
\end{tabular}
\caption{(a) Worst-user rate versus $N$, (b) rate loss of LR-QAOA versus $S$, and (c) execution time versus $N$ on \texttt{ibm\_fez} quantum processor.}
\label{fig:rates}
\end{figure*}

Fig.~\ref{fig:rates}(a) shows the worst-user rate versus $N$ for $M=3$ and $S=1024$, averaged over 1000 instances. It can be observed that the performance of LR-QAOA almost overlaps the exhaustive search upper bound for all $N$, even though it looks at only a tiny fraction of the patterns, whereas SDR falls slightly further behind as $N$ grows. Furthermore, the classical greedy activation and PGGA methods show a much larger performance gap since both of these methods add one PA at a time and stop as soon as no single addition improves $\Gamma$. Because every new PA takes power away from the others, adding one PA often lowers $\Gamma$, even when adding several PAs together would raise it. Therefore, these baseline methods stop too early, with fewer active PAs than the optimal pattern.




We define the rate loss of a method that returns the pattern $\hat{\mathbf a}$ as
\begin{equation}
\mathcal L(\hat{\mathbf a})=\log_2\!\big(1+\bar\gamma\,\Gamma(\mathbf a^{\rm opt})\big)-\log_2\!\big(1+\bar\gamma\,\Gamma(\hat{\mathbf a})\big),
\label{eq:loss}
\end{equation}
where $\bar\gamma=P\eta/\sigma_w^2$ and $\mathbf a^{\rm opt}$ is the solution of \eqref{eq:P}, found by exhaustive search over all $2^N-1$ patterns.
Fig.~\ref{fig:rates}(b) shows the rate loss of LR-QAOA versus $S$ for $N=12$, 16, and 20, using post-selection only. It can be observed that for $M=1$, a few tens of patterns $S$ are enough to bring the loss below 0.03\,bps/Hz (dotted line), whereas $M=3$ needs a few hundred, and more as $N$ grows. Hence, more users need more evaluated patterns to reach the same rate loss.


\subsection{Quantum Hardware Results}
\label{sec:hw}
Next, we execute LR-QAOA on the 156-qubit IBM Heron processor \texttt{ibm\_fez} for $N=6–120$, with $p=5$, $M=3$, and $32$ shots per instance, using Pauli twirling and dynamical decoupling. Given QPU access costs, we limit sampling to $32$ shots and refine the best measured shot with a short local search of at most $512$ evaluations. SDR uses the same number of randomized candidates. The QPU time excludes queuing and transpilation overheads,
whereas exhaustive search and SDR run on a CPU. Since exhaustive search cannot be evaluated beyond $N=24$ on a classical CPU, the runtime results are extrapolated beyond $N=24$. Exhaustive search and SDR runtimes are reported for a server with two AMD EPYC 9555 64-core processors
(128 cores, 256 threads) and 1\,TB of RAM, where SDR is solved by CVXPY \cite{diamond2016cvxpy}.


Fig.~\ref{fig:rates}(c) compares the runtimes of the proposed LR-QAOA on actual quantum hardware with that of the classical exhaustive search and SDR on classical CPU.
It can be observed that the QPU run-time of LR-QAOA stays at 2\,s for every $N$, because it is set by the fixed overhead of each job rather than by the number of qubits. Exhaustive search, on the other hand, doubles its CPU time with every added PA and becomes slower than LR-QAOA beyond $N\approx26$. Furthermore, it can be observed that the runtime of SDR keeps growing with $N$ and is more than that of the quantum  runtime for beyond $N=24$ PAs.

Table~\ref{tab:hw} compares the run times and worst user rates of the LR-QAOA on quantum hardware against the classical exhaustive search (Exh.) and SDR-based methods. Here, Shots denotes the worst-user rate of the best of the 32 measured patterns, i.e., LR-QAOA with post-selection on $\Gamma$ but without local search. For $N\ge32$, the time of exhaustive search is extrapolated from the 36.4\,ns per pattern measured at $N=24$. The table shows that LR-QAOA keeps its advantage over SDR on real hardware as well. It finds the optimum for every $N\le20$ and comes very close at $N=24$, while SDR misses it at several sizes. For larger $N$, the two methods are nearly equal at $N=32$, and LR-QAOA is clearly better at $N=64$ and 120. The LR-QAOA based on shots without post selection, however, are far from optimal, because the circuit contains $pN(N-1)/2$ two-qubit blocks whose errors add up even with error mitigation. Most of the gain therefore comes from the local search post-selection. Therefore, we do not claim a quantum advantage on today's hardware. However, our quantum hardware results show that the QPU run-time does not grow with $N$, so the advantage over exhaustive search will become usable as quantum hardware gate errors come down in the future.

\begin{table}[!t]
\centering
\caption{Runtime and Worst-User Rate Comparison for $M=3$ Users.
}
\label{tab:hw}
\footnotesize
\setlength{\tabcolsep}{2.5pt}
\begin{tabular}{@{}r|ccc|cccc@{}}
\toprule
 & \multicolumn{3}{c|}{Time (s)} & \multicolumn{4}{c}{Rate (bps/Hz)}\\
$N$ & Exh. & QPU & SDR & Exh. & LR-QAOA & Shots & SDR\\
\midrule
6   & $9.9\cdot10^{-5}$ & 2.0 & 0.07 & 4.137 & 4.137 & 3.915 & 4.137\\
8   & $9.5\cdot10^{-5}$ & 2.0 & 0.34 & 4.402 & 4.402 & 4.402 & 4.402\\
10  & $1.8\cdot10^{-4}$ & 2.0 & 0.07 & 4.460 & 4.460 & 3.941 & 4.058\\
12  & $4.6\cdot10^{-4}$ & 2.0 & 0.53 & 4.677 & 4.677 & 4.001 & 4.677\\
16  & $7.0\cdot10^{-3}$ & 2.0 & 0.83 & 5.213 & 5.213 & 4.016 & 5.213\\
20  & $5.5\cdot10^{-2}$ & 2.0 & 0.90 & 5.167 & 5.167 & 4.439 & 4.973\\
24  & 0.61 & 2.0 & 2.65 & 5.737 & 5.712 & 4.057 & 5.550\\
32  & $1.6\cdot10^{2}$ & 2.0 & 47.6 & -- & 5.585 & 3.605 & 5.603\\
64  & $6.7\cdot10^{11}$ & 2.0 & 83.8 & -- & 6.330 & 3.769 & 5.918\\
120 & $4.8\cdot10^{28}$ & 2.0 & 212.8 & -- & 7.343 & 3.874 & 3.182\\
\bottomrule
\end{tabular}
\end{table}

\section{Conclusion}
\label{sec:conclusion}
We have proposed a training-free quantum optimization algorithm that selects which pinching antennas to activate in PASS to maximize the worst-user rate. We mapped the max-min SNR optimization problem to a two-local Ising Hamiltonian with one qubit per PA and no penalty terms through a quadratic surrogate that is lower-bounded by the worst-user SNR, with equality for a single user. Since this bound loosens for several users, the LR-QAOA samples are post-selected on the exact objective and refined by a short local search. Our simulation results show that LR-QAOA outperforms greedy activation, PGGA, and SDR-based classical baseline methods when all methods test the same number
of PA on/off patterns. Furthermore, on the IBM quantum processor \texttt{ibm\_fez}, the LR-QAOA takes only 2\,s of QPU run-time for up to 120 qubits, and with local search post-selection, LR-QAOA finds the optimum up to $N=20$ and beats SDR for large $N$. However, in the NISQ era, due to quantum hardware gate noise, most of the gain of the proposed LR-QAOA comes from the local search post-selection. Future work will focus on investigating shallower hardware-aware circuits, stronger error mitigation, and multi-waveguide PASS systems.

\bibliographystyle{IEEEtran}
\bibliography{ref}
\end{document}